\documentclass[preprint, preprintnumbers ,amsmath,amssymb]{revtex4}
\usepackage{amssymb}
\usepackage{amsmath}
\usepackage{graphicx}
\usepackage{multirow}

\begin{document}



\title{Properties of the Schr{\"o}dinger Theory of Electrons in
Electromagnetic Fields}



\author{Viraht Sahni $^{1}$ and Xiao-Yin Pan $^{2}$}

\affiliation{$^{1}$ Brooklyn College and The Graduate School of the
City University of
New York, New York, New York 10016.\\
$^{2}$ Department of Physics, Ningbo University, Ningbo, 315211,
China.}


\date{\today}

\begin{abstract}
The Schr{\"o}dinger theory of electrons in an external
electromagnetic field can be described from the perspective of the
individual electron via the `Quantal Newtonian' laws (or
differential virial theorems). These laws are in terms of
`classical' fields whose sources are quantal expectations of
Hermitian operators taken with respect to the wave function.   The
laws reveal the following physics: (a) In addition to the external
field, each electron experiences an internal field whose components
are representative of a specific property of the system such as the
correlations due to the Pauli exclusion principle and Coulomb
repulsion, the electron density, kinetic effects, and an internal
magnetic field component. (The response of the electron is described
by the current density field.); (b) The scalar potential energy of
an electron is the work done in a conservative field which is the
sum of the internal and Lorentz fields.  It is thus inherently
related to the properties of the system.  Its constituent
property-related components are hence known.   It is a known
functional of the wave function; (c) As such the Hamiltonian is a
functional of the wave function, thereby revealing the intrinsic
self-consistent nature of the Schr{\"o}dinger equation.  This then
provides a path for the determination of the exact wave function.
(d) With the Schr{\"o}dinger equation written in self-consistent
form, the Hamiltonian now admits via the Lorentz field a new term
that explicitly involves the external magnetic field.  The new
understandings are explicated for the stationary state case by
application to a quantum dot in a magnetostatic field in both a
ground and excited state.  For the time-dependent case, the same
states of the quantum dot in both a magnetostatic and a
time-dependent electric field are considered.
\end{abstract}


\maketitle



\section{Introduction}

In this paper we explain new understandings of \cite{1}
Schr{\"o}dinger theory of the electronic structure of matter, and of
the interaction of matter with external static and time-dependent
electromagnetic fields. Matter -- atoms, molecules, solids, quantum
wells, two-dimensional electron systems such as at semiconductor
heterojunctions, etc., -- is defined here as a system of $N$
electrons in an external electrostatic field ${\boldsymbol{\cal{E}}}
({\bf{r}})  = - {\boldsymbol{\nabla}} v ({\bf{r}})$ where $v
({\bf{r}})$ is the scalar potential energy of an electron. The added
presence of a magnetostatic field ${\boldsymbol{\cal{B}}} ({\bf{r}})
= {\boldsymbol{\nabla}} \times {\bf{A}} ({\bf{r}})$, with ${\bf{A}}
({\bf{r}})$ the vector potential, corresponds to the Zeeman, Hall,
Quantum Hall, and magneto-caloric effects, magnetoresistance,
quantum dots, etc. The interaction of radiation with matter such as
laser-atom interactions, photo-electric effects at metal surfaces,
etc., are described by the case of external time-dependent
electromagnetic fields.   The insights are arrived at by describing
Schr{\"o}dinger theory from the perspective \cite{2,3} of the
\emph{individual} electron.  This perspective is arrived at via the
`Quantal Newtonian' second law \cite{4,5,6,7} (or the time-dependent
differential virial theorem) for each electron, with the first law
\cite{8} being a description of stationary-state theory. The laws
are a description of the system \cite{2,3} in terms of `classical'
fields whose sources are quantal in that they are expectations of
Hermitian operators taken with respect to the wave function.  This
manner of depiction makes the description of Schr{\"o}dinger theory
tangible in the classical sense.  The new understandings described
are a consequence of these `Quantal Newtonian' laws.

A principal insight into Schr{\"o}dinger theory arrived at is that
the Schr{\"o}dinger equation can be written in self-consistent form.
To explain what we mean, consider first the stationary-state case.
It is proved via the `Quantal Newtonian' first law, that the
Hamiltonian $\hat{H}$ for the system of electrons in a static
electromagnetic field is a functional of the wave function $\Psi$,
\emph{i.e.} $\hat{H} = \hat{H} [\Psi]$.  Hence, the corresponding
Schr{\"o}dinger equation can be written as $\hat{H} [\Psi] \Psi = E
\Psi$. Thus, the eigenfunctions $\Psi$ and eigenenergies $E$ of the
Schr{\"o}dinger equation can be obtained self-consistently. This
form of eigenvalue equation is mathematically akin to that of
Hartree-Fock and Hartree theories in which the corresponding
Hamiltonian $\hat{H}^{\mathrm{HF}}$ is a functional of the single
particle orbitals $\phi_{i}$ of the Slater determinant wave
function. The corresponding integro-differential eigenvalue
equations are then $\hat{H}^{\mathrm{HF}} [\phi_{i}] \phi_{i} =
\epsilon_{i} \phi_{i}$ .  The orbitals $\phi_{i}$ and the
eigenenergies $\epsilon_{i}$ are obtained by self-consistent
solution of the equation \cite{9, 10}. There are many other
formalisms whereby the solution is obtained self-consistently such
as, for example, the Optimized Potential Method \cite{11, 12} and
the Hartree and Pauli-correlated approximations within Quantal
density functional theory \cite{13, 14}. (In general, eigenvalue
equations of the form $\hat{\bf{L}} [\zeta] \zeta = \lambda \zeta$
are solved in an iterative self-consistent manner.) In the
time-dependent case, it is shown via the `Quantal Newtonian' second
law that the Hamiltonian $\hat{H} (t) = \hat{H} [\Psi (t)]$, so that
the self-consistent form of the Schr{\"o}dinger equation is $\hat{H}
[\Psi (t)] \Psi (t) = i
\partial \Psi (t)/\partial t$.

Other understandings achieved show that the scalar potential energy
of an electron $v ({\bf{r}})$ is the work done in a conservative
field ${\boldsymbol{\cal{F}}} ({\bf{r}})$. The components of this
field are separately representative of properties of the system such
as the correlations due to the Pauli exclusion principle and Coulomb
repulsion, the electron density, kinetic effects, an internal
magnetic field contribution, and the Lorentz field.  The constituent
property-related components of the potential $v ({\bf{r}})$ are thus
known. The components of the field ${\boldsymbol{\cal{F}}}
({\bf{r}})$ are expectation values of Hermitian operators taken with
respect to the wave function $\Psi$.  Thus, the potential $v
({\bf{r}})$, (and hence the Hamiltonian), is a known functional of
the wave function. Finally, the presence of the Lorentz field in the
expression for $v ({\bf{r}})$, admits a term involving the magnetic
field ${\boldsymbol{\cal{B}}} ({\bf{r}})$ in the Schr{\"o}dinger
equation as written in self-consistent form. These insights all lead
to a fundamentally different way of thinking of the Schr{\"o}dinger
equation.

The new physics is explicated for the stationary-state case by
application to the ground and first excited singlet state of a
two-dimensional quantum dot in a magnetostatic field. For the
time-dependent case, the same states of the quantum dot in a
magnetostatic field perturbed by a time-dependent electric field are
considered.

We begin with a brief summary of the manner in which Schr{\"o}dinger
theory is presently understood and practiced.  For this consider
stationary-state theory for a system of $N$ electrons in an external
electrostatic field ${\boldsymbol{\cal{E}}} ({\bf{r}})  = -
{\boldsymbol{\nabla}} v ({\bf{r}})$ and magnetostatic field
${\boldsymbol{\cal{B}}} ({\bf{r}}) = {\boldsymbol{\nabla}} \times
{\bf{A}} ({\bf{r}})$. The Schr{\"o}dinger equation in atomic units
(charge of electron $-e, |e|=\hbar=m=1$) together with the
assumption of $c=1$ is
\begin{equation}
\bigg[ \frac{1}{2} \sum_{i} \big( \hat{\bf{p}}_{i} + {\bf{A}}
({\bf{r}}_{i}) \big)^{2} + \frac{1}{2} \sideset{}{'}\sum_{i,j}
\frac{1}{|{\bf{r}}_{i} - {\bf{r}}_{j}|} + \sum_{i} v ({\bf{r}}_{i})
\bigg] \Psi ({\bf{X}}) = E \Psi ({\bf{X}}),
\end{equation}
where the terms of the Hamiltonian are the physical kinetic,
electron-interaction potential, and scalar potential energy
operators; $\{ \Psi ({\bf{X}}); E \}$ the eigenfunctions and
eigenvalues; ${\bf{X}} = {\bf{x}}_{1}, {\bf{x}}_{2}, \ldots,
{\bf{x}}_{N}$ ; ${\bf{x}} = {\bf{r}} \sigma$ ; $({\bf{r}} \sigma)$
the spatial and spin coordinates.

We note the following salient features of the above Schr{\"o}dinger
equation:\\
(a) As a consequence of the correspondence principle, it is the
vector potential ${\bf{A}} ({\bf{r}})$ and not the magnetic field
${\boldsymbol{\cal{B}}} ({\bf{r}})$ that appears in it.  This fact
is significant, and is expressly employed to explain, for example,
the Bohm-Aharonov \cite{15} effect in which a vector potential can
exist in a region of no magnetic field.  The magnetic field
${\boldsymbol{\cal{B}}} ({\bf{r}})$ appears in the
Schr{\"o}dinger equation only following the choice of gauge;\\
(b)  The characteristics of the potential energy operator $v
({\bf{r}})$ are the following:

\par
(i)  For the $N$-electron system, it is assumed that the canonical
kinetic and electron-interaction potential energy operators are
known.  As such, the potential $v ({\bf{r}})$ is considered an
\emph{extrinsic} input to the Hamiltonian.

\par
(ii)  The potential energy function $v ({\bf{r}})$ is \emph{assumed}
known, e.g. it could be Coulombic, harmonic, Yukawa, etc.

\par
(iii)  By assumption, the potential $v ({\bf{r}})$ is
\emph{path-independent}.
\newline
With the Hamiltonian \emph{known},
the Schr{\"o}dinger differential equation is then solved for $\{
\Psi ({\bf{X}}); E \}$. Physical observables are determined as
expectations of Hermitian operators taken with respect to $\Psi
({\bf{X}})$.

We initially focus on the stationary-state case.  In Sect. II, we
briefly describe the single-electron perspective of time-independent
Schr{\"o}dinger theory via the `Quantal Newtonian' first law.  The
explanation of the new understandings achieved is given in Sect.
III.  These ideas are further elucidated in Sect. IV by the example
of a quantum dot in a magnetostatic field.  Both a ground and
excited state are considered.  The extension to the time-dependent
case via the `Quantal Newtonian' second law is discussed in Sect. V.
Concluding remarks are made in Sect. VI together with a comparison
of the self-consistent method and the variational and
constrained-search variational methods for the determination of the
wave function.

\section{Stationary State Theory: `Quantal Newtonian' First Law}

In order to better understand the `Quantal Newtonian' laws for each
electron, we first draw a parallel to Newton's laws for the
individual particle.  Hence, consider a system of $N$ classical
particles that obey Newton's third law, exert forces on each other
that are equal and opposite, directed along the line joining them,
and are subject to an external force.  Then Newton's second law for
the $i^{th}$ particle is
\begin{equation}
{\bf{F}}^{\mathrm{ext}}_{i} + \sideset{}{'}\sum_{j} {\bf{F}}_{j i} =
d {\bf{p}}_{i}/d t,
\end{equation}
where ${\bf{F}}^{\mathrm{ext}}_{i}$ is the external force,
${\bf{F}}_{j i}$ the \emph{internal} force on the $i^{th}$ particle
due to the $j^{th}$ particle, and ${\bf{p}}_{i}$ the linear momentum
response of the $i^{th}$ particle to these forces.  In summing Eq.
(2) over all the particles, the internal force contribution
vanishes, leading to Newton's second law.

Newton's first law for the $i^{th}$ particle is
\begin{equation}
{\bf{F}}^{\mathrm{ext}}_{i} + \sideset{}{'}\sum_{j} {\bf{F}}_{j i} =
0.
\end{equation}
Again, on summing over all the particles, the internal force
component vanishes leading to Newton's first law.

The `Quantal Newtonian' first law for the quantum system described
by Eq. (1) -- (the counterpart to Newton's first law for each
particle) -- states that the sum of the external
${\boldsymbol{\cal{F}}}^{\mathrm{ext}} ({\bf{r}})$ and
\emph{internal} ${\boldsymbol{\cal{F}}}^{\mathrm{int}} ({\bf{r}})$
fields experienced by each electron vanish \cite{2,3,16,17}:
\begin{equation}
{\boldsymbol{\cal{F}}}^{\mathrm{ext}} ({\bf{r}}) +
{\boldsymbol{\cal{F}}}^{\mathrm{int}} ({\bf{r}}) = 0.
\end{equation}
The law is valid for arbitrary gauge and derived employing the
continuity condition ${\boldsymbol{\nabla}} \cdot {\bf{j}}
({\bf{r}}) =0$.  Here ${\bf{j}} ({\bf{r}})$ is the physical current
density which is the expectation $< \Psi ({\bf{X}}) | \hat{\bf{j}}
({\bf{r}}) | \Psi ({\bf{X}}) >$ with the operator $\hat{\bf{j}}
({\bf{r}}) = \{ \frac{1}{2 i} \sum_{k} \big[
{\boldsymbol{\nabla}}_{{\bf{r}}_{k}} \delta ({\bf{r}}_{k} -
{\bf{r}}) + \delta ({\bf{r}}_{k} - {\bf{r}})
{\boldsymbol{\nabla}}_{{\bf{r}}_{k}} \big] + \hat{\rho} ({\bf{r}})
{\bf{A}} ({\bf{r}}) \}$ and $\hat{\rho} ({\bf{r}}) = \sum_{k} \delta
({\bf{r}}_{k} - {\bf{r}})$, the density operator.  The external
field is the sum of the electrostatic ${\boldsymbol{\cal{E}}}
({\bf{r}})$ and Lorentz ${\boldsymbol{\cal{L}}} ({\bf{r}})$ fields
\cite{16}:
\begin{equation}
{\boldsymbol{\cal{F}}}^{\mathrm{ext}} ({\bf{r}}) =
{\boldsymbol{\cal{E}}} ({\bf{r}}) - {\boldsymbol{\cal{L}}}
({\bf{r}}) = - {\boldsymbol{\nabla}} v ({\bf{r}}) -
{\boldsymbol{\cal{L}}} ({\bf{r}}),
\end{equation}
where ${\boldsymbol{\cal{L}}} ({\bf{r}})$ is defined in terms of the
Lorentz `force' ${\boldsymbol{\ell}} ({\bf{r}})$ as
${\boldsymbol{\cal{L}}} ({\bf{r}}) = {\boldsymbol{\ell}}
({\bf{r}})/\rho ({\bf{r}})$, with $\rho ({\bf{r}}) = < \Psi
({\bf{X}}) | \hat{\rho} ({\bf{r}})| \Psi ({\bf{X}}) >$ is the
density, and where ${\boldsymbol{\ell}} ({\bf{r}}) = {\bf{j}}
({\bf{r}}) \times {\boldsymbol{\cal{B}}} ({\bf{r}})$.

The internal field ${\boldsymbol{\cal{F}}}^{\mathrm{int}}
({\bf{r}})$ is the sum of the electron-interaction
${\boldsymbol{\cal{E}}}_{\mathrm{ee}} ({\bf{r}})$, kinetic
${\boldsymbol{\cal{Z}}} ({\bf{r}})$, differential density
${\boldsymbol{\cal{D}}} ({\bf{r}})$, and internal magnetic
${\boldsymbol{\cal{I}}} ({\bf{r}})$ fields \cite{16}:
\begin{equation}
{\boldsymbol{\cal{F}}}^{\mathrm{int}} ({\bf{r}}) =
{\boldsymbol{\cal{E}}}_{\mathrm{ee}} ({\bf{r}}) -
{\boldsymbol{\cal{Z}}} ({\bf{r}}) - {\boldsymbol{\cal{D}}}
({\bf{r}}) - {\boldsymbol{\cal{I}}} ({\bf{r}}).
\end{equation}
These fields are defined in terms of the corresponding `forces'
${\bf{e}}_{\mathrm{ee}} ({\bf{r}})$, ${\bf{z}} ({\bf{r}})$,
${\bf{d}} ({\bf{r}})$, and ${\bf{i}} ({\bf{r}})$. (Each `force'
divided by the (charge) density $\rho ({\bf{r}})$ constitutes the
corresponding field.)  The `force' ${\bf{e}}_{\mathrm{ee}}
({\bf{r}})$, representative of electron correlations due to the
Pauli exclusion principle and Coulomb repulsion, is obtained via
Coulomb's law via its quantal source, the pair-correlation function
$P ({\bf{r r}}'): {\bf{e}}_{\mathrm{ee}} ({\bf{r}}) = \int d
{\bf{r}}' P ({\bf{r r}}') ({\bf{r - r}}')/|{\bf{r - r}}'|^{3}$, with
$P ({\bf{r r}}')$ the expectation of the pair operator $\hat{P}
({\bf{r r}}') = \sum'_{i,j} \delta ({\bf{r}}_{i} - {\bf{r}})\delta
({\bf{r}}_{j} - {\bf{r}})$; the kinetic `force' ${\bf{z}}
({\bf{r}})$, representative of kinetic effects, is obtained from its
quantal source, the single-particle density matrix $\gamma ({\bf{r
r}}'):z_{\alpha} ({\bf{r}}) = 2 \sum_{\beta} \nabla_{\beta}
t_{\alpha \beta} ({\bf{r}})$, where the kinetic energy tensor
$t_{\alpha \beta} ({\bf{r}}) = (1/4) [\partial^{2}/\partial
r'_{\alpha}
\partial r''_{\beta} +
\partial^{2}/\partial r'_{\beta}
\partial r''_{\alpha}] \gamma ({\bf{r}}' {\bf{r}}'')|_{{\bf{r}}' =
{\bf{r}}'' = r}$ with $\gamma ({\bf{r r'}})$ the expectation of the
operator $\hat{\gamma} ({\bf{r r'}}) = \hat{A} + i \hat{B}$,
$\hat{A} = \frac{1}{2} \sum_{j} [\delta ({\bf{r}}_{j} - {\bf{r}})
T_{j} ({\bf{a}}) + \delta ({\bf{r}}_{j} - {\bf{r}}') T_{j}
(-{\bf{a}})]$, $\hat{B} = - \frac{i}{2} \sum_{j} [\delta
({\bf{r}}_{j} - {\bf{r}}) T_{j} ({\bf{a}}) - \delta ({\bf{r}}_{j} -
{\bf{r}}') T_{j} (-{\bf{a}})]$, with $T_{j} ({\bf{a}})$ a
translation operator such that $T_{j} ({\bf{a}}) \psi ( \ldots
{\bf{r}}_{j} \ldots) = \psi ( \ldots {\bf{r}}_{j} + {\bf{a}},
\ldots)$; the differential density `force', representative of the
density is ${\bf{d}} ({\bf{r}}) = - \frac{1}{4}
{\boldsymbol{\nabla}} \nabla^{2} \rho ({\bf{r}})$, the quantal
source being the density $\rho ({\bf{r}})$; and internal magnetic
`force' ${\bf{i}} ({\bf{r}})$ whose quantal source is the current
density ${\bf{j}} ({\bf{r}}):i_{\alpha} ({\bf{r}}) = \sum_{\beta}
\nabla_{\beta} I_{\alpha \beta} ({\bf{r}})$, $I_{\alpha \beta}
({\bf{r}}) = [j_{\alpha} ({\bf{r}}) A_{\beta} ({\bf{r}}) + j_{\beta}
({\bf{r}}) A_{\alpha} ({\bf{r}})] - \rho ({\bf{r}}) A_{\alpha}
({\bf{r}}) A_{\beta} ({\bf{r}})$.  The components of the total
energy $E$ -- the kinetic, electron-interaction, internal magnetic,
and external -- can each be expressed in integral virial form in
terms of the respective fields \cite{16}.  For example, the
electron-interaction energy $E_{\mathrm{ee}} = \int \rho ({\bf{r}})
{\bf{r}} \cdot {\boldsymbol{\cal{E}}}_{\mathrm{ee}} ({\bf{r}}) d
{\bf{r}}$, the kinetic energy $T = - \frac{1}{2} \int \rho
({\bf{r}}) {\bf{r}} \cdot {\boldsymbol{\cal{Z}}} ({\bf{r}}) d
{\bf{r}}$, etc.

\section{New Understandings}

We next discuss the new insights achieved via the single-electron
perspective.  They are valid for both ground and excited states.

(i) In addition to the external electrostatic
${\boldsymbol{\cal{E}}} ({\bf{r}})$ and Lorentz
${\boldsymbol{\cal{L}}} ({\bf{r}})$ fields, each electron
experiences an internal field ${\boldsymbol{\cal{F}}}^{\mathrm{int}}
({\bf{r}})$. This field via its
${\boldsymbol{\cal{E}}}_{\mathrm{ee}} ({\bf{r}})$ component is
representative not only of Coulomb correlations as one might expect,
but also those due to the Pauli exclusion principle due to the
antisymmetric nature of the wave function. Additionally there is a
component ${\boldsymbol{\cal{Z}}} ({\bf{r}})$ representative of the
motion of the electrons; a component ${\boldsymbol{\cal{D}}}
({\bf{r}})$ representing the density, a fundamental property of the
system \cite{18,19}; and a term ${\boldsymbol{\cal{I}}} ({\bf{r}})$
that arises as a consequence of the external magnetic field
\cite{16}. \emph{Hence, each electron experiences an internal field
that encapsulates all the basic properties of the system.}  As in
classical physics, in summing over all the electrons, the
contribution of the internal field vanishes, leading thereby to
Ehrenfest's (first law) theorem: $\int \rho ({\bf{r}})
{\boldsymbol{\cal{F}}}^{\mathrm{ext}} ({\bf{r}}) d {\bf{r}} =0$. (In
fact, each component of the internal field is shown to separately
vanish.)

(ii)    The `Quantal Newtonian' first law Eq. (4) affords a rigorous
physical interpretation of the external electrostatic potential $ v
({\bf{r}})$: \emph{It is the work done to move an electron from some
reference point at infinity to its position at ${\bf{r}}$ in the
force of a conservative field ${\boldsymbol{\cal{F}}} ({\bf{r}})$:}
\begin{equation}
v ({\bf{r}}) = \int^{\bf{r}}_{\infty} {\boldsymbol{\cal{F}}}
({\bf{r}}') \cdot d {\boldsymbol{\ell}}',
\end{equation}
where ${\boldsymbol{\cal{F}}} ({\bf{r}}) =
{\boldsymbol{\cal{F}}}^{\mathrm{int}} ({\bf{r}}) -
{\boldsymbol{\cal{L}}} ({\bf{r}}) =
{\boldsymbol{\cal{E}}}_{\mathrm{ee}} ({\bf{r}}) -
{\boldsymbol{\cal{Z}}} ({\bf{r}}) - {\boldsymbol{\cal{D}}}
({\bf{r}}) - {\boldsymbol{\cal{I}}} ({\bf{r}}) -
{\boldsymbol{\cal{L}}} ({\bf{r}})$. Since $\nabla \times
{\boldsymbol{\cal{F}}} ({\bf{r}}) = 0$, this work done is
\emph{path-independent}. Thus, we now understand, in the rigorous
classical sense of a potential being the work in a conservative
field, that $v ({\bf{r}})$ represents a potential energy \emph{viz}.
that of an electron.

(iii)   What the physical interpretation of the potential $v
({\bf{r}})$ further shows is that it can no longer be thought of as
an \emph{independent} entity.  It is \emph{intrinsically} dependent
upon \emph{all the properties of the system} via the various
components of the internal field
${\boldsymbol{\cal{F}}}^{\mathrm{int}} ({\bf{r}})$, and the Lorentz
${\boldsymbol{\cal{L}}} ({\bf{r}})$ field through the current
density ${\bf{j}} ({\bf{r}})$.  Hence, the potential energy function
$v ({\bf{r}})$ is comprised of the sum of constituent functions each
representative of a property of the system.

(iv) As each component of the internal field
${\boldsymbol{\cal{F}}}^{\mathrm{int}} ({\bf{r}})$ (and the Lorentz
field ${\boldsymbol{\cal{L}}} ({\bf{r}})$) are obtained from quantal
sources that are expectations of Hermitian operators taken with
respect to the wave function $\Psi ({\bf{X}})$, we see that the
field ${\boldsymbol{\cal{F}}} ({\bf{r}})$ is a functional of $\Psi
({\bf{X}})$, \emph{i.e}. ${\boldsymbol{\cal{F}}} ({\bf{r}}) =
{\boldsymbol{\cal{F}}}  [ \Psi ({\bf{X}})]$.  Thus, (from Eq. (7)),
$v ({\bf{r}})$ \emph{is a functional of} $\Psi ({\bf{X}}): v
({\bf{r}}) = v[\Psi ({\bf{X}})]$. \emph{The functional} $v[\Psi
({\bf{X}})]$ \emph{is exactly known} [via Eq. (7)].

(v) On substituting the functional $v[\Psi ({\bf{X}})]$ into Eq.
(1), the Schr{\"o}dinger equation may then be written as
\begin{equation}
\bigg[ \frac{1}{2} \sum_{i} (\hat{\bf{p}}_{i} + {\bf{A}}
({\bf{r}}_{i}))^{2} + \frac{1}{2} \sideset{}{'}\sum_{i,j}
\frac{1}{|{\bf{r}}_{i} - {\bf{r}}_{j}|} + \sum_{i} v[\Psi]
({\bf{r}}_{i})\bigg] \Psi ({\bf{X}}) = E \Psi ({\bf{X}}).
\end{equation}
or equivalently as
\begin{equation}
\bigg[ \frac{1}{2} \sum_{i} (\hat{\bf{p}}_{i} + {\bf{A}}
({\bf{r}}_{i}))^{2} + \frac{1}{2} \sideset{}{'}\sum_{i,j}
\frac{1}{|{\bf{r}}_{i} - {\bf{r}}_{j}|} + \sum_{i}
\int_{\infty}^{{\bf{r}}_{i}} {\boldsymbol{\cal{F}}} [\Psi]
({\bf{r}}) \cdot d {\boldsymbol{\ell}} \bigg] \Psi ({\bf{X}}) = E
\Psi ({\bf{X}}).
\end{equation}
In general, with the Hamiltonian a functional of $\Psi ({\bf{X}})$,
the Schr{\"o}dinger equation can be written as $\hat{H} [\Psi] \Psi
({\bf{X}}) = E \Psi ({\bf{X}})$.  \emph{In this manner, the
intrinsic self-consistent nature of the Schr{\"o}dinger equation
becomes evident.}  (Recall that what is meant by the functional $v
[\Psi]$ is that for each different $\Psi ({\bf{X}})$ one obtains a
different $v [\Psi] ({\bf{r}})$.)  To solve the equation (see Fig.
1), one begins with an approximation to $\Psi ({\bf{X}})$.  With
this approximate $\Psi ({\bf{X}})$ one determines the various
quantal sources and the fields
${\boldsymbol{\cal{F}}}^{\mathrm{int}} ({\bf{r}})$ and
${\boldsymbol{\cal{L}}} ({\bf{r}})$ (for an external
${\boldsymbol{\cal{B}}} ({\bf{r}})$), and the work done in the sum
of these fields.  One then solves the integro-differential equation
to determine a new approximate solution $\Psi ({\bf{X}})$ and
eigenenergy $E$.  This $\Psi ({\bf{X}})$, in turn, will lead to a
new $v ({\bf{r}})$  (via Eq. (7)), and by solution of the equation
to a new $\Psi ({\bf{X}})$ and $E$. The process is continued till
the input $\Psi ({\bf{X}})$ to determine $v [\Psi]$ leads to the
same output $\Psi ({\bf{X}})$ on solution of the equation or
equivalently till self-consistency is achieved.  The exact $\hat{H}
[\Psi]$, $\Psi ({\bf{X}})$, $E$ are obtained in the final iteration
of the self-consistency procedure.

\begin{figure}
\hspace*{-4cm}
\includegraphics[bb= 00 00 200 330 width=1\textwidth]{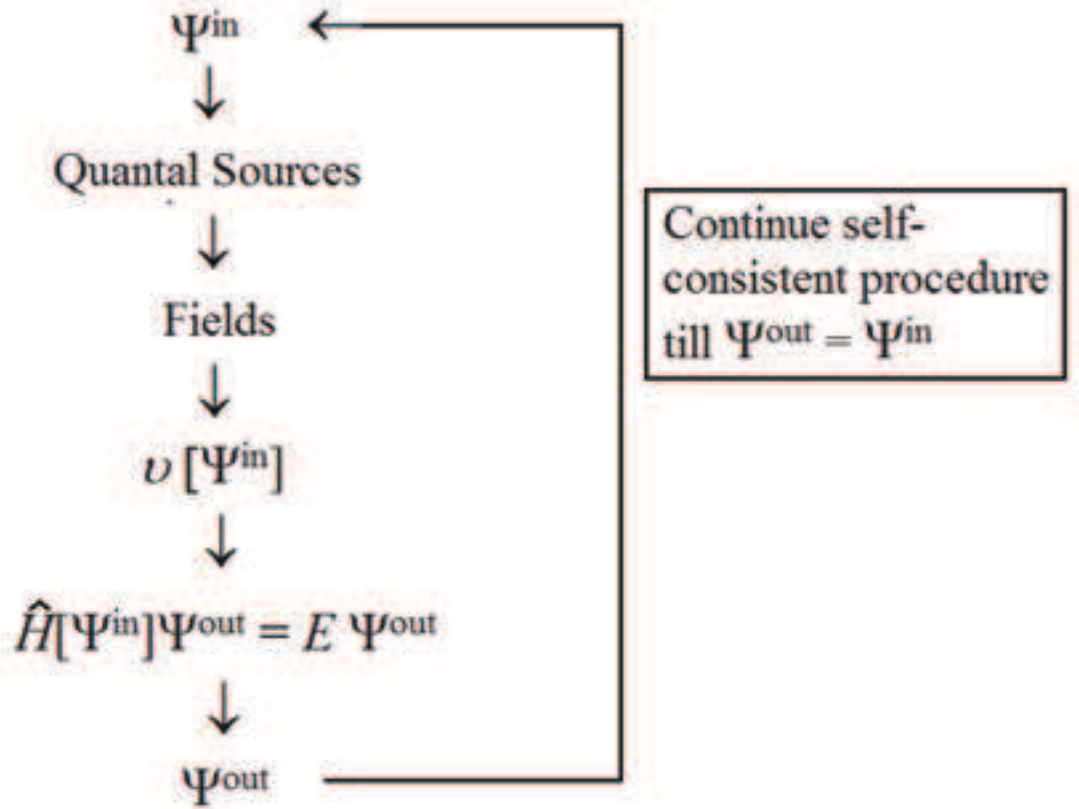}
\caption{Procedure for the self-consistent solution of the
Schr{\"o}dinger equation.}
\end{figure}

In any self-consistent procedure, different external potentials $v
({\bf{r}})$ can be obtained based on the choice of the initial
approximate input wave function $\Psi ({\bf{X}})$.  In atoms,
molecules or solids, the potential $v ({\bf{r}})$ obtained
self-consistently would be Coulombic.  In quantum dots it would be
harmonic, and so on.  One must begin with an educated accurate guess
\emph{apropos} to the physical system of interest for the initial
input.  Otherwise one may not achieve self-consistency. Thus, for
example, in self-consistent quantal density functional theory
calculations on atoms \cite{3,13,14}, the initial input wave
function for an atom is the solution of the prior atom of the
Periodic Table. In general, for any self-consistent calculation, it
is only after self-consistency is achieved that one must judge and
test whether the solution is physically meaningful. (Note that in
this manner, the external potential $v ({\bf{r}})$ and hence the
Hamiltonian is determined self-consistently.)

In principle, the above procedure is mathematically entirely akin to
the fully-self-consistent solution of the integro-differential
equations of Hartree \cite{9} and Hartree-Fock \cite{10} theories,
the Optimized Potential method \cite{11,12}, Quantal density
functional theory \cite{3}, etc.   In each of these cases, the
corresponding integro-differential equations are of the form
$\hat{H} [\zeta_{i}] \zeta_{i} ({\bf{x}}) = \lambda_{i} \zeta_{i}
({\bf{x}})$, where $\hat{H}$ is the corresponding Hamiltonian and
$\zeta_{i}, \lambda_{i}$ the single particle orbitals and
eigenvalues, respectively. This eigenvalue equation is of the same
form as that of the Schr{\"o}dinger equation written in
self-consistent form but with the generalization to the
many-electron system.  Thus, we now understand that the
Schr{\"o}dinger equation too can be thought of as being a
self-consistent equation.  This perspective of Schr{\"o}dinger
theory is new.

(We note that there exists a `Quantal Newtonian' first law for
Hartree, Hartree-Fock, and local effective potential theories
\cite{2,3}. Hence, the external potential $v ({\bf{r}})$ of these
theories can also be expressed as the work done in a conservative
field, and thus replaced in the corresponding equations by a known
functional of the requisite Slater determinant.)

(vi)   \emph{Observe that in writing the Schr{\"o}dinger equation as
in Eqs. (8), (9), the magnetic field ${\boldsymbol{\cal{B}}}
({\bf{r}})$ now appears in the Hamiltonian explicitly via the
Lorentz field ${\boldsymbol{\cal{L}}} ({\bf{r}})$.} (See Eq. (7).)
It is the intrinsic self-consistent nature of the equation that
demands the presence of ${\boldsymbol{\cal{B}}} ({\bf{r}})$ in the
Hamiltonian.  In other words, as the Hamiltonian  $\hat{H} [\Psi]$
is being determined self-consistently, all the information of the
physical system -- electrons and fields -- must be incorporated in
it.  (Of course, equivalently the field ${\boldsymbol{\cal{B}}}
({\bf{r}})$ could be expressed in terms of the vector potential
${\bf{A}} ({\bf{r}})$. This then shows that when written in
self-consistent form, there exists another component of the
Hamiltonian involving the vector potential.)

(vii)  The presence of a solely electrostatic external field
${\boldsymbol{\cal{E}}} ({\bf{r}}) = - {\boldsymbol{\nabla}} v
({\bf{r}})$ is a special case of the stationary state theory
discussed above. This case then constitutes the description of
matter as defined previously.

\section{Example of a Quantum Dot}
To explicate the new physics of stationary-state Schr{\"o}dinger
theory, we consider a ground and first excited singlet state of a
two electron, two-dimensional quantum dot in an external
magnetostatic field \cite{20,21}.  The external scalar potential in
the Hamiltonian of Eq. (1) is then $v ({\bf{r}}) = \frac{1}{2}
\omega^{2}_{0} r^{2}$, with $\omega_{0}$ the harmonic frequency. The
ground $\psi_{0} ({\bf{r}}_{1} {\bf{r}}_{2})$ \cite{16} and excited
$\psi_{1} ({\bf{r}}_{1} {\bf{r}}_{2})$ \cite{22} state wave
functions of the quantum dot in the symmetric gauge ${\bf{A}}
({\bf{r}}) = \frac{1}{2} {\boldsymbol{\cal{B}}} ({\bf{r}}) \times
{\bf{r}}$, are respectively,
\begin{equation}
\psi_{0} ({\bf{r}}_{1} {\bf{r}}_{2}) = C_{0} e^{- \Omega(R^{2} +
\frac{1}{4} r^{2})} (1 + r),
\end{equation}
and
\begin{eqnarray}
\psi_{1} ({\bf{r}}_{1} {\bf{r}}_{2}) = C_{1} e^{- \Omega(R^{2} +
\frac{1}{4} r^{2})} \bigg[1 &+& r + \bigg(\frac{\Omega} {4} -
0.436815 \bigg) r^{2}
\nonumber \\
&+& \bigg(\frac{\Omega} {4} - 0.353786 \bigg) r^{3}\bigg],
\end{eqnarray}
where ${\bf{R}} = ({\bf{r}}_{1} + {\bf{r}}_{2})/2$, $r =
|{\bf{r}}_{1} - {\bf{r}}_{2}|$, $C_{0} = \Omega^{\frac{3}{2}}/ \pi
[2 + \Omega + \sqrt{2 \pi \Omega}]^{\frac{1}{2}}$, $C_{1} =
0.108563$, $\Omega = \sqrt{k_{\mathrm{eff}}}$, the effective force
constant $k_{\mathrm{eff}} = \omega_{0}^{2} + \omega_{L}^{2} = 1$
for the ground state, and $k_{\mathrm{eff}} = 0.471716$ for the
excited case, with $\omega_{L} = B/2$ the Larmor frequency, $E_{0} =
3.000000$ a.u., $E_{1} = 3.434076$ a.u..

In Figs. 2 and 3 we plot the corresponding electron-interaction
${\boldsymbol{\cal{E}}}_{\mathrm{ee}} ({\bf{r}})$, kinetic
${\boldsymbol{\cal{Z}}} ({\bf{r}})$, and differential density
${\boldsymbol{\cal{D}}} ({\bf{r}})$ components of the internal
${\boldsymbol{\cal{F}}}^{\mathrm{int}} ({\bf{r}})$ field.  The total
magnetic field contribution is incorporated into the effective force
constant $k_{\mathrm{eff}}$.  Observe that $-
{\boldsymbol{\cal{E}}}_{\mathrm{ee}} ({\bf{r}}) +
{\boldsymbol{\cal{Z}}} ({\bf{r}}) + {\boldsymbol{\cal{D}}}
({\bf{r}}) = - k_{\mathrm{eff}} r$.  This then demonstrates the
satisfaction of the `Quantal Newtonian' first law of Eq. (4).

The example of the quantum dot above can be thought of as being the
final iteration of the self-consistent procedure in which the exact
potential $v ({\bf{r}})$, wave function $\Psi$, and energy $E$ are
obtained.  To see this, consider the initial choice of wave
functions to be the following:
\begin{equation}
\Psi_{0} ({\bf{r}}_{1} {\bf{r}}_{2}) = C_{0} e^{- \Omega_{0} (R^{2}
+ \frac{1}{4} r^{2})} (1 + a_{0} r),
\end{equation}
and
\begin{equation}
\Psi_{1} ({\bf{r}}_{1} {\bf{r}}_{2}) = C_{1} e^{- \Omega_{1} (R^{2}
+ \frac{1}{4} r^{2})} (1 + a_{1} r + b_{1} r^{2} + c_{1} r^{3}),
\end{equation}
where $C_{0}, C_{1}, \Omega_{0}, \Omega_{1}, a_{0}, a_{1}, b_{1},
c_{1}$ are constants.  Let us next assume that for some random
iteration, the values of these coefficients turn out to be $C_{0} =
0.135646$, $\Omega_{0} = 1.000000$, $a_{0} = 1.000000$; $C_{1} =
0.108563$, $\Omega_{1} = 0.686816$, $a_{1} = 1.000000$, $b_{1} = -
0.265111$, $c_{1} = -0.182082$.  One then determines the various
fields from the corresponding wave functions and plots them.  On
adding the fields ${\boldsymbol{\cal{D}}} ({\bf{r}})$ and
${\boldsymbol{\cal{Z}}} ({\bf{r}})$ one obtains the dot-dash lines
as shown in Figs. 2 and 3 for (${\boldsymbol{\cal{D}}} ({\bf{r}}) +
{\boldsymbol{\cal{Z}}} ({\bf{r}})$).  Adding $-
{\boldsymbol{\cal{E}}}_{\mathrm{ee}} ({\bf{r}})$ to these lines, one
then obtains a straight (dashed) line $- k_{\mathrm{eff}} r$ in each
case. On substituting this $- k_{\mathrm{eff}} r$ back into the
Schr{\"o}dinger equation and solving, one obtains the same wave
functions $\Psi_{0}, \Psi_{1}$ and energies $E_{0}, E_{1}$ as that
of Eqs. (10) and (11). Additionally, it becomes clear that the
potential $v ({\bf{r}})$ is harmonic.  This then constitutes the
final iteration of the self-consistency procedure.

\begin{figure}
\includegraphics[bb=50 50 800 930,  width=1\textwidth]{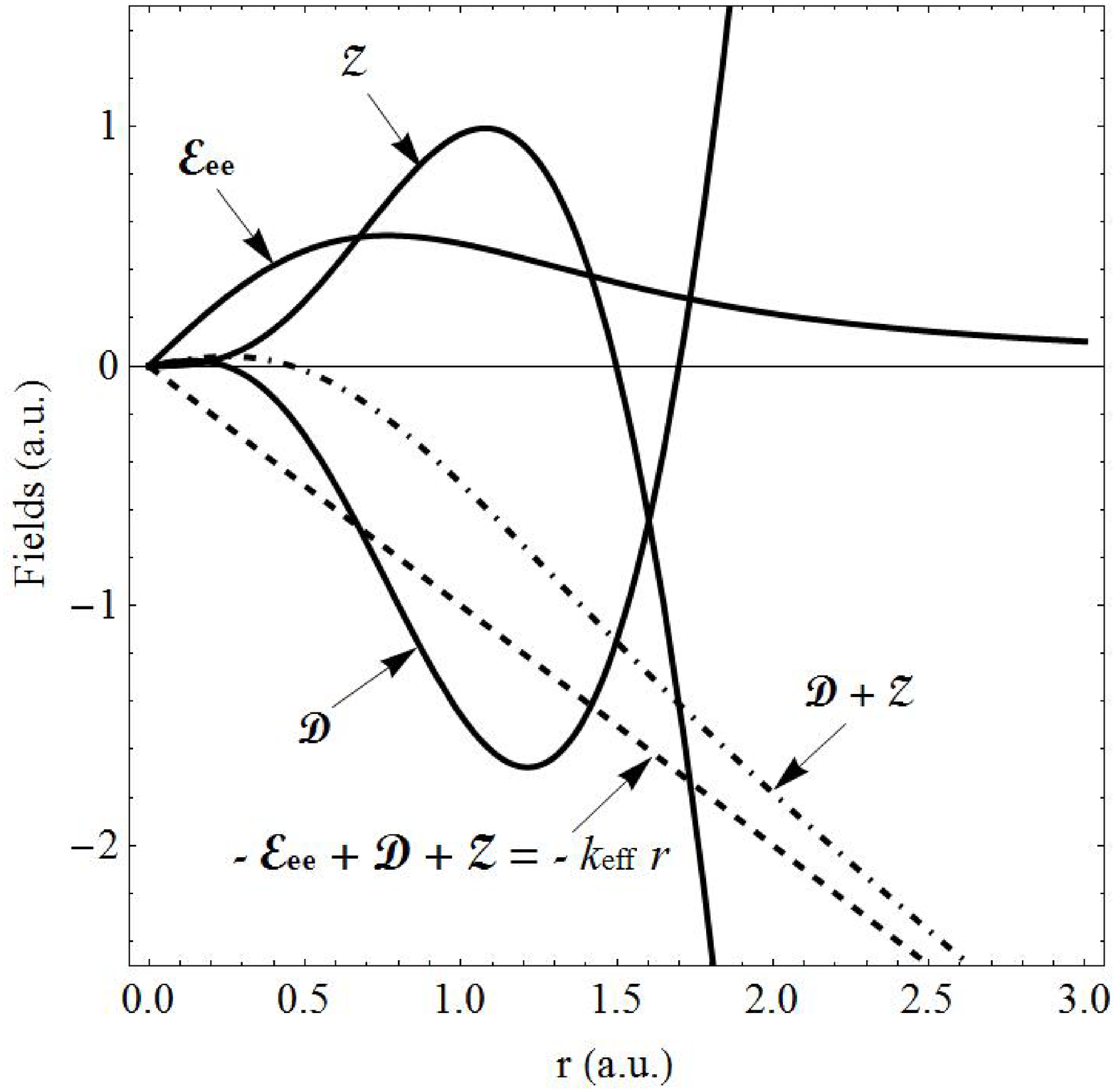}
\caption{The electron-interaction
${\boldsymbol{\cal{E}}}_{\mathrm{ee}} ({\bf{r}})$, kinetic
${\boldsymbol{\cal{Z}}} ({\bf{r}})$, and differential density
${\boldsymbol{\cal{D}}} ({\bf{r}})$ components of the internal field
${\boldsymbol{\cal{F}}}^{\mathrm{int}} ({\bf{r}})$ for the ground
state of a quantum dot in a magnetic field.  The sums
${\boldsymbol{\cal{D}}} ({\bf{r}}) + {\boldsymbol{\cal{Z}}}
({\bf{r}})$, and $-{\boldsymbol{\cal{E}}}_{\mathrm{ee}} ({\bf{r}}) +
{\boldsymbol{\cal{Z}}} ({\bf{r}}) + {\boldsymbol{\cal{D}}}
({\bf{r}}) = - k_{\mathrm{eff}} r$ with $k_{\mathrm{eff}} =1$ are
also plotted.}
\end{figure}

\begin{figure}
\includegraphics[bb=00 00 750 830,  width=1\textwidth]{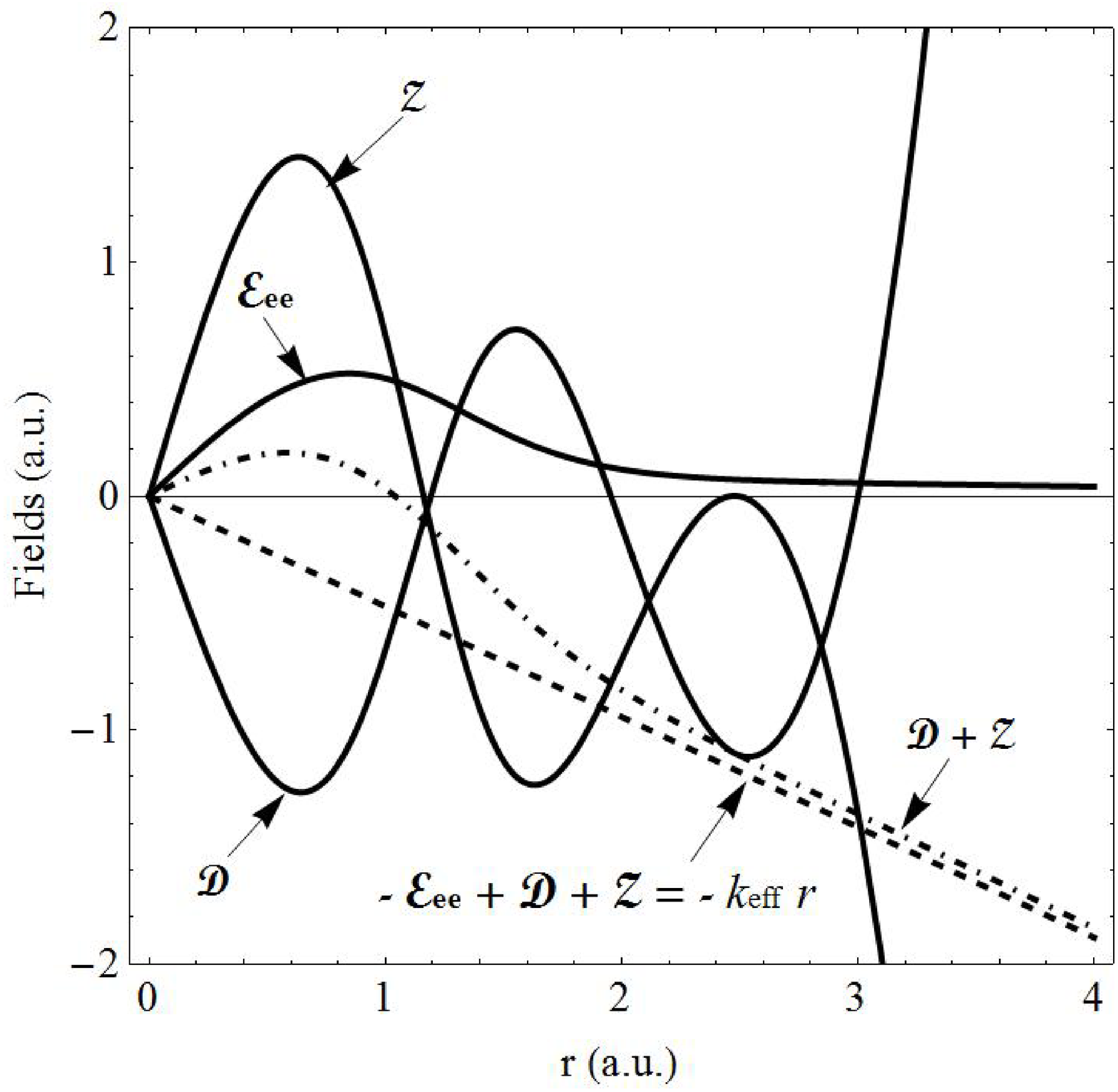}
\caption{Same as in Fig. 2 but for the first excited singlet state
with $k_{\mathrm{eff}} = 0.471716$.}
\end{figure}

\section{Time-dependent Theory: `Quantal Newtonian' Second Law}

The above conclusions are generalizable to the TD case by
considering the external field to be
${\boldsymbol{\cal{F}}}^{\mathrm{ext}} ({\bf{r}} t) =
{\boldsymbol{\cal{E}}} ({\bf{r}} t) = - {\boldsymbol{\nabla}} v
({\bf{r}} t)$. In this case, the `Quantal Newtonian' second law for
each electron -- the quantal equivalent to Newton's second law of
Eq. (2) -- is \cite{2,4,5}
\begin{equation}
{\boldsymbol{\cal{F}}}^{\mathrm{ext}} ({\bf{r}} t) +
{\boldsymbol{\cal{F}}}^{\mathrm{int}} ({\bf{r}} t) =
{\boldsymbol{\cal{J}}} ({\bf{r}} t),
\end{equation}
where ${\boldsymbol{\cal{F}}}^{\mathrm{int}} ({\bf{r}} t)$ is given
by the TD version of Eq. (6) (without the ${\boldsymbol{\cal{I}}}
({\bf{r}} t)$ term) and the response of the electron is described by
the current density field ${\boldsymbol{\cal{J}}} ({\bf{r}} t)  =
(1/\rho ({\bf{r}} t))
\partial {\bf{j}} ({\bf{r}} t)/\partial t$. The corresponding
external potential energy $v [\Psi] ({\bf{r}} t)$ functional of the
wave function $\Psi ({\bf{X}} t)$ is then the work done at each
instant of time in a conservative field:
\begin{equation}
v [\Psi] ({\bf{r}} t) = \int^{\bf{r}}_{\infty}
{\boldsymbol{\cal{F}}} ({\bf{r}}' t) \cdot d {\boldsymbol{\ell}}',
\end{equation}
where ${\boldsymbol{\cal{F}}} ({\bf{r}} t) =
{\boldsymbol{\cal{F}}}^{\mathrm{int}} ({\bf{r}} t) -
{\boldsymbol{\cal{J}}} ({\bf{r}} t)$, with the TD self-consistent
Schr{\"o}dinger equation being
\begin{equation}
\bigg[ \frac{1}{2} \sum_{i} \hat{\bf{p}}_{i}^{2} + \frac{1}{2}
\sideset{}{'}\sum_{i,j} \frac{1}{|{\bf{r}}_{i} - {\bf{r}}_{j}|} +
\sum_{i} v[\Psi] ({\bf{r}}_{i} t)\bigg] \Psi ({\bf{X}} t) = i
\frac{\partial \Psi ({\bf{X}} t)} {\partial t}.
\end{equation}
Since ${\boldsymbol{\nabla}} \times {\boldsymbol{\cal{F}}} ({\bf{r}}
t) =0$, the work done $v ({\bf{r}} t)$ at each instant of time is
\emph{path-independent} and thus a potential energy. Again, on
summing Eq, (14) over all the electrons, the contribution of
${\boldsymbol{\cal{F}}}^{\mathrm{int}} ({\bf{r}} t)$ vanishes,
leading to Ehrenfest's (second law) theorem $\int \rho ({\bf{r}} t)
[{\boldsymbol{\cal{F}}}^{\mathrm{ext}} ({\bf{r}} t) -
{\boldsymbol{\cal{J}}} ({\bf{r}} t)]d {\bf{r}} = 0$. The further
generalization of the `Quantal Newtonian' second law to the case of
an external TD electromagnetic field with ${\boldsymbol{\cal{E}}}
({\bf{r}})  = - {\boldsymbol{\nabla}} v ({\bf{r}})$, ${\bf{E}}
({\bf{r}} t) = - {\boldsymbol{\nabla}} \phi ({\bf{r}} t) -
\partial {\bf{A}} ({\bf{r}} t)/\partial t$, ${\boldsymbol{\cal{B}}}
({\bf{r}} t)= {\boldsymbol{\nabla}} \times {\bf{A}} ({\bf{r}} t)$,
is given in \cite{7}.

As an example of the insights for the time-dependent case, consider
the two-dimensional two-electron quantum dot in an external
magnetostatic field ${\boldsymbol{\cal{B}}} ({\bf{r}}) =
{\boldsymbol{\nabla}} \times  {\bf{A}} ({\bf{r}})$ perturbed by a
time-dependent electric field ${\bf{E}} (t)$.  The wave function of
this system \cite{23, 24}, known as the Generalized Kohn Theorem, is
comprised of a phase factor times the unperturbed wave function in
which the coordinates of each electron are translated by a
time-dependent function that satisfies the classical equation of
motion.  Hence, if the unperturbed wave function is known, the time
evolution of all properties is known. As the wave functions for a
ground and excited state of the unperturbed quantum dot are given by
Eqs. (10), (11), the corresponding solutions of the time-dependent
Schr{\"o}dinger equation, and therefore of all the various fields,
is obtained.    At the initial time, $t = 0$, the results are those
of Figs. 2 and 3. Observables that are expectations of
non-differential operators such as the density $\rho ({\bf{r}} t)$,
the electron-interaction field
${\boldsymbol{\cal{E}}}_{\mathrm{ee}}({\bf{r}} t)$, etc., are simply
the time-independent functions shifted in time.

\section{Concluding Remarks}

In conclusion, we have arrived at new insights into the
Schr{\"o}dinger theory of electrons in electromagnetic fields via
the `Quantal Newtonian' first and second laws for each electron.  A
principal understanding is that the scalar potential energy of an
electron $\{ v ({\bf{r}})/ v ({\bf{r}} t) \}$ is a known functional
of the wave function $\{ \Psi ({\bf{X}})/ \Psi ({\bf{X}} t) \}$.  As
such the Hamiltonian $\{ \hat{H}/\hat{H} (t) \}$ is a functional of
the wave function: $\{ \hat{H} [\Psi ({\bf{X}})]/\hat{H} [\Psi
({\bf{X}}) t] \}$. Thus the Schr{\"o}dinger equation can now be
thought of as one whose solution can be obtained self-consistently.
A path for the determination of the exact wave function is thus
formulated. Such a path is feasible given the advent of present-day
high computing power.  A second understanding achieved is that it is
now possible to write the scalar potential $[v ({\bf{r}})/v
({\bf{r}} t)]$ as the sum of component functions each of which is
representative of a specific property of the system such as the
correlations due to the Pauli exclusion principle and Coulomb
repulsion, kinetic and magnetic effects, and the electron density.
Such a property-related division of the scalar potential is shown by
the example of the quantum dot in a magnetostatic field given in the
text.  Another interesting observation is that in its
self-consistent form, in addition to the vector potential, which
appears in the Schr{\"o}dinger equation as a consequence of the
correspondence principle, the magnetic field now too appears in the
equation because of the `Quantal Newtonian' laws.  \emph{Ex post
facto}, we now understand that this must be the case as the
Hamiltonian itself is being determined self-consistently.

It is interesting to compare the self-consistent method for the
determination of the wave function in the stationary ground state
case to that of the variational method. The latter is associated
principally with the property of the total energy.  An approximate
parametrized variational wave function correct to $O (\delta)$ leads
to an upper bound for the energy that is correct to $O
(\delta^{2})$.  Such a wave function is accurate in the region where
the principal contribution to the energy arises.  However, all other
observables obtained as the expectation of Hermitian single- and
two-particle operators are correct only to the same order as that of
the wave function, \emph{viz}. to $O (\delta)$.  A better
approximate variational wave function is one that leads to a lower
value of the energy.  There is no guarantee that other observables
representative of different regions of configuration space are
thereby more accurate. On the other hand, in the self-consistent
procedure, achieved say to a desired accuracy of five decimal
places, \emph{all} the properties are correct to the same degree of
accuracy.  An improved wave function would be one correct to a
greater decimal accuracy.  As a point of note, the
constrained-search variational method \cite{25,26,27} expands the
variational space of approximate parametrized wave functions by
considering the wave function $\Psi$ to be a functional of a
function $\chi$, \emph{i.e.} $\Psi = \Psi [\chi]$. One searches over
all functions $\chi$ such that the wave function $\Psi [\chi]$ is
normalized, gives the exact (theoretical or experimental) value of
an observable, while leading to a rigorous upper bound to the
energy. In this manner, the wave function functional $\Psi [\chi]$
is accurate not only in the region contributing to the energy, but
also that of the observable.  The self-consistent solution of the
Schr{\"o}dinger equation, however, is accurate to the degree
required, throughout configuration space.

We do not address here the broader procedural aspects of the
self-consistency, nor the implications of the explicit presence of
the magnetic field in it. These issues constitute current and future
research.

The authors acknowledge Lou Massa and Marlina Slamet for their
critique of the paper. VS is supported in part by the Research
Foundation of the City University of New York.  XP is supported by
the National Natural Science Foundation of China (Grant No:
11275100).

\end{document}